\documentclass[11pt,twoside]{article}
\usepackage{asp2010}
\newcommand{{\bfv}}{\mbox{\boldmath$v$\unboldmath}}
\newcommand{{\bfa}}{\mbox{\boldmath$a$\unboldmath}}
\newcommand{{\bfmu}}{\mbox{\boldmath$\mu$\unboldmath}}

\resetcounters

\begin{document}

\title{Pulsars are born as magnetars}
\author{Ricardo Heras
\affil{Preparatoria Abierta, SEIEM Toluca Edo.~de M\'exico}}

\begin{abstract}
This paper suggests the idea that all neutron stars experienced at birth an ultrafast decay of their magnetic fields from their initial values to their current surface values. If the electromagnetic energy radiated during this field decay is converted into kinetic energy of the neutron star via the radiation reaction mechanism then the decay time is of the order of $10^{-4}$s provided that the initial magnetic fields lie in the range of $10^{14}\!-\!10^{16}$G. This means that all neutron stars are born with magnetic fields typical of magnetars. According to this model the neutron stars acquire their observed high space velocities during the birth ultrafast decay of their magnetic fields. The origin of this field decay points to magnetic instabilities occurring at the end of the birth process.
\end{abstract}

\section{Introduction}
The neutron stars PSR J2144$-$3933 and SGR 0418+5729 have strongly called attention because their respective periods $P\!=\!8.5$ s and $P\!=\!9.1$ s are typical of magnetars but their respective magnetic fields $B_s\!= \!1.9\!\times \!10^{12}$G and $B_s\!=\!7.5\!\times \!10^{12}$G are typical of radio pulsars (Tiengo et al 2011; Rea et al. 2010). This seems to indicate that both pulsars were born as magnetars, but their magnetic fields decayed to intensities typical of radio pulsars. This paper suggests that this decay was ultrafast and occurred during the birth of these stars. More generally, this paper suggests the idea that all neutron stars at birth experienced an abrupt exponential decay of their magnetic fields from their initial values $B_0$ to their current surface values $B_s$. The nascent neutron stars then radiated electromagnetic energy during the specific time $\tau_s$.
If this radiative energy is converted into kinetic energy then the time $\tau_s$ is shown to be of the order of $10^{-4}$s provided that the initial magnetic fields lie in the range of $10^{14}\!-\!10^{16}$G. This means that neutron stars are born with magnetic fields of magnetars. Magnetic instabilities occurring at the end of the birth process [Geppert \& Rheinhardt(2011)] seems to be the cause of the field decay. The basic assumption in this paper is that there is a radiation reaction process that can convert all radiative energy produced by the field decay into kinetic energy.
Neutron stars acquire their high velocities during the decay of their magnetic fields.

\section{Birth ultrafast magnetic field decay of neutron stars}

The Larmor formula for the power radiated by a time-varying magnetic dipole moment $P=2\ddot{\mu}^2/(3c^3)$  and the estimate
$\ddot{\mu}\!\sim\!\mu_0/\tau^2$, where $\tau$ is the characteristic time in the exponential field decay law $B(t)\!=\!B_0e^{-t/\tau},$ imply the equation
$P\!\simeq \!2\mu_0^2/(3c^3\tau^4),$ which can be used together  with the relation $\mu_0\!=\!B_0R^3/2$ to yield the power radiated by a neutron star of radius $R$ and an initial magnetic field $B_0$: $P\!\simeq \!B_0^2 R^6/~(6c^3\tau^4).$
Consider now the specific time $\tau_s$ elapsed during the field decay from $B_0$ to the current surface magnetic field $B_s$. The condition $B(\tau_s)\!=\!B_s$ and the exponential law $B(t)\!=\!B_0e^{-t/\tau}$ imply $B_s\!=\!B_0e^{-\tau_s/\tau}$ and thus $\tau_s\!=\!\tau\ln(B_0/B_s).$ During the time $\tau_s$ the energy radiated $E_{\rm rad}\!\simeq \tau_s P$ is given by $E_{\rm rad}\!\simeq \!B_0^2 R^6\ln(B_0/B_s)/(6c^3\tau^3).$
If this energy is converted into kinetic energy $ E_{\rm kin}=Mv^2/2$
 of the neutron star then $B_0^2 R^6\ln(B_0/B_s)/(6c^3\tau^3)=Mv^2/2$, which implies $\tau= [B_0^2R^6\ln(B_0/B_s)/(3 c^3 Mv^2)]^{1/3}.$
To get the order of magnitude of $\tau$ consider a neutron star with $B_s\!=\!10^{12}$G and $v=400$ km/s. If one assumes $B_0\!=\!1.4\!\times\!10^{15}$G then $\tau\!\approx\! 3.3 \!\times\! 10^{-5}{\rm s}\! \approx \!R/c.$ This is the theoretical shortest time for accelerating the neutron star. One can adopt the time $\tau=R/c$ as the characteristic time in the law $B(t)\!=\!B_0e^{-t/\tau}$. This adoption implies the specific time $\tau_s= (R /c)\ln(B_0/B_s).$ For the abrupt field decays one has $2.3  \leq\ln(B_0/B_s)\leq4.4$ and then $
\tau_s\!\sim\!10^{-4}$s, which indicates that the decay from $B_0$ to $B_s$ is ultrafast if $B_0$ lies in the range of $10^{14}\!\!-\!\!10^{16}$G.
Using $\tau=R/c$ the assumed energy conversion reads $B_0^2 R^3\ln(B_0/B_s)/6=Mv^2/2$. This equation and $v\!\approx\!\sqrt{3/2}\:v_{\perp}$, $M\!= \!1.4 M_\odot$ and $R\!=\!10$ km implies the following expression the initial magnetic field:
\begin{equation}
B_0= B_s\:e^{W(\sigma v_{\perp}^2B_s^{-2})/2},
\end{equation}
where $\sigma\!=\!2.52\! \times \!10^{16}$gr/cm$^{3}$ and $W(x)$ is the Lambert function, which is defined as the inverse of the function $f(x)\!=\!xe^{x}$ satisfying $W(x)e^{W(x)}\!=\!x$. Remarkably, Eq.~(1) predicts the value of $B_0$ in terms of the current values for $ v_{\perp}$ and $B_s$ of the neutron stars.

\section{Applications}
As already mentioned, the enigmatic pulsar PSR J2144-3933 exhibits magnetar properties but its magnetic field is typical of radio pulsars:
$B_s\!=\!1.9\times\! 10^{12}$G.  This pulsar has the transverse velocity $v_{\perp}\!=\!135$ km/s (Tiengo et al. 2011). In this case
Eq.~(4) predicts $B_0\!\approx \! 6.29\times \!10^{14}$G indicating that this pulsar was born with a magnetic field typical of magnetars. Analogously, the SGR 0418+5729 also exhibits magnetar properties but it has a field of radio pulsar: $B_s\!\simeq \!7.5\!\times \!10^{12}$G (Rea et al. 2010). The velocity $v_{\perp}$ of this pulsar has not been reported yet. It is reasonable to assume that this velocity lies in the range $100\!-\!400$ km/s. If this would be the case, then Eq.~(1) predicts that the initial magnetic field SGR 0418+5729 lies in the range $5.4\times \!10^{14}\!-\!1.9\times \!10^{15}$G indicating that this SGR was born with an initial magnetic field typical of magnetars.

Equation~(1) can be applied to the set of 130 neutron stars whose periods $P_s$ are in the interval: $.02$ s$ \leq P_s\!\!< \!\!1$ s. The values of $P_s, B_s$ and $v_{\perp}$ are taken from the ATNF Pulsar Catallogue V 1.43 (www.atnf.csiro.au/research/pulsar/psrcat; Manchester et al. 2005)]. The average values are $\widetilde{B_s}=1.28\times 10^{12}$G and $\widetilde{v_{\perp}}=409.06$ km/s. For this set of pulsars, Eq.~(1) predicts $\widetilde{B_0}=1.7\times 10^{15}$G. One can also consider the case of the 9 isolated millisecond pulsars in the
interval: $.0015$ s $\! \leq P_s<\! .009$ s. The average values are $\widetilde{B_s}=2.59\times 10^{8}$G and $\widetilde{v_{\perp}}\!=\!60.11$ km/s. For this set, Eq.~(1) predicts $\widetilde{B_0}\!=\!1.8\times 10^{14}$G.
One concludes then that all neutron star are born with magnetic fields of magnetars.

\end{document}